\documentclass[conference,twocolumn,10pts]{IEEEtran}

\usepackage{graphicx} 
\usepackage{amsmath, amsthm, amssymb, graphicx,amsfonts, multirow}
\title{Gaussian Two-Way Channels With\\ Discrete Inputs and Quantized Outputs}
\author{
\IEEEauthorblockN{Ershad Banijamali}
\IEEEauthorblockA{School of Computer Science\\University of Waterloo \\
Email: sbanijam@uwaterloo.ca}
}
\begin{document}
\maketitle
\newcounter{MYtempeqncnt}
\begin{abstract}
In this paper\footnote{\textit{Accepted in Biennial Symposium on Communications, BSC2016}}, Gaussian two-way channel with uniform output quantization is studied.
For Gaussian inputs, the optimum uniform finite-level quantizer is determined
numerically for different values of Signal-to-Noise Ratio (SNR). The two-way channel with constellation-based transmitters is then investigated. A formulation for the so-called Shannon achievable region of this channel is developed and numerical computations of this region are presented for particular constellations. It is shown that if one transmitter utilizes a rotated version of the constellation used at the other transmitter, the Shannon achievable region can be enlarged.

\end{abstract}
\begin{section}{Introduction and Preliminaries}
\vspace{-0cm}
The two-way channel \cite{sh}, in its conventional form, consists of  two nodes or users. Each node has its own transmitter and receiver. The nodes intend to transmit their messages to each other over one single channel. A fundamental feature of this channel is the so-called self-interference, i.e.,  the leakage of one node's transmitted signal at its own receiver. That is, the desired signal for one user plays the role of interference for the other user.

The capacity region of a two-way channel in its general form is still unknown. In \cite{sh}, Shannon established inner and outer bounds on the capacity region of a two-way channel.  Suppose $X_i$ and $\tilde{Y}_i$ represent the transmitted signal and received signal for $i^{th}$ node, $i \in \{1,2\}$. Lets denote the rate of the code-book that carries information from transmitter 1 to receiver 2 by $R_{1}$ and from transmitter 2 to receiver 1 by $R_{2}$. The outer bound includes all pairs of $(R_1,R_2)$ satisfying the inequalities
\begin{equation}
\begin{array}{cc}
\label{eqn: rate}
R_1\leq I(X_1;\tilde{Y}_2|X_2),\\ 
R_2\leq I(X_2;\tilde{Y}_1| X_1),
\end{array}
\end{equation}
where $X_1$ and $X_2$ have an arbitrary joint distribution. As for the inner bound, these expressions still hold, however, $X_1$ and $X_2$ are
independent random variables.

Fig. \ref{fig: channel} illustrates the so-called Gaussian Two-Way Channel (GTWC) given by
\begin{equation}
\begin{array}{cc}
      \tilde{Y}_1=aX_1+bX_2+Z_2,\\  
      \tilde{Y}_2=cX_1+dX_2+Z_1,
      \end{array}
      \end{equation}
where $X_1$ and $X_2$ represent the transmitted signals with power constraints $E\{|X_{i}|^{2}\}\leq P_i$ for $i=1,2$ and $Z_1$ and $Z_2$ are additive noises at the receiver sides. Moreover, $Z_1$ and $Z_2$ are independent zero-mean Gaussian random variables, $N(0,\sigma_z^2)$. Since the physical distance from one node's receiver to its own transmitter is usually much smaller than the distance to the other node's transmitter, the interfering signal has much higher power than the desired signal, i.e., $d$ and $a$ are much larger than $c$ and $b$, respectively\footnote{Using RF techniques, one may considerably reduce self-interference \cite{Dr}.}. In [2], the capacity region of GTWC is shown to be the rectangular region
\begin{equation}
\label{r1}
\begin{array}{cc}
      R_{1}\leq\frac{1}{2}\log\left(1+\frac{c^{2}P_{1}}{\sigma_z^2}\right),  \\\\
      R_{2}\leq\frac{1}{2}\log\left(1+\frac{b^{2}P_{2}}{\sigma_z^2}\right).
\end{array}
\end{equation}

From (\ref{r1}) it can be seen that the capacity achieving inputs are Gaussian and each side can completely cancel the self-interference. As such, GTWC is equivalent to two orthogonal (parallel) Gaussian point-to-point
channels.

\begin{figure}[!t]
\centering
\includegraphics[trim = 0mm 0mm 0mm 5mm,width=80mm]{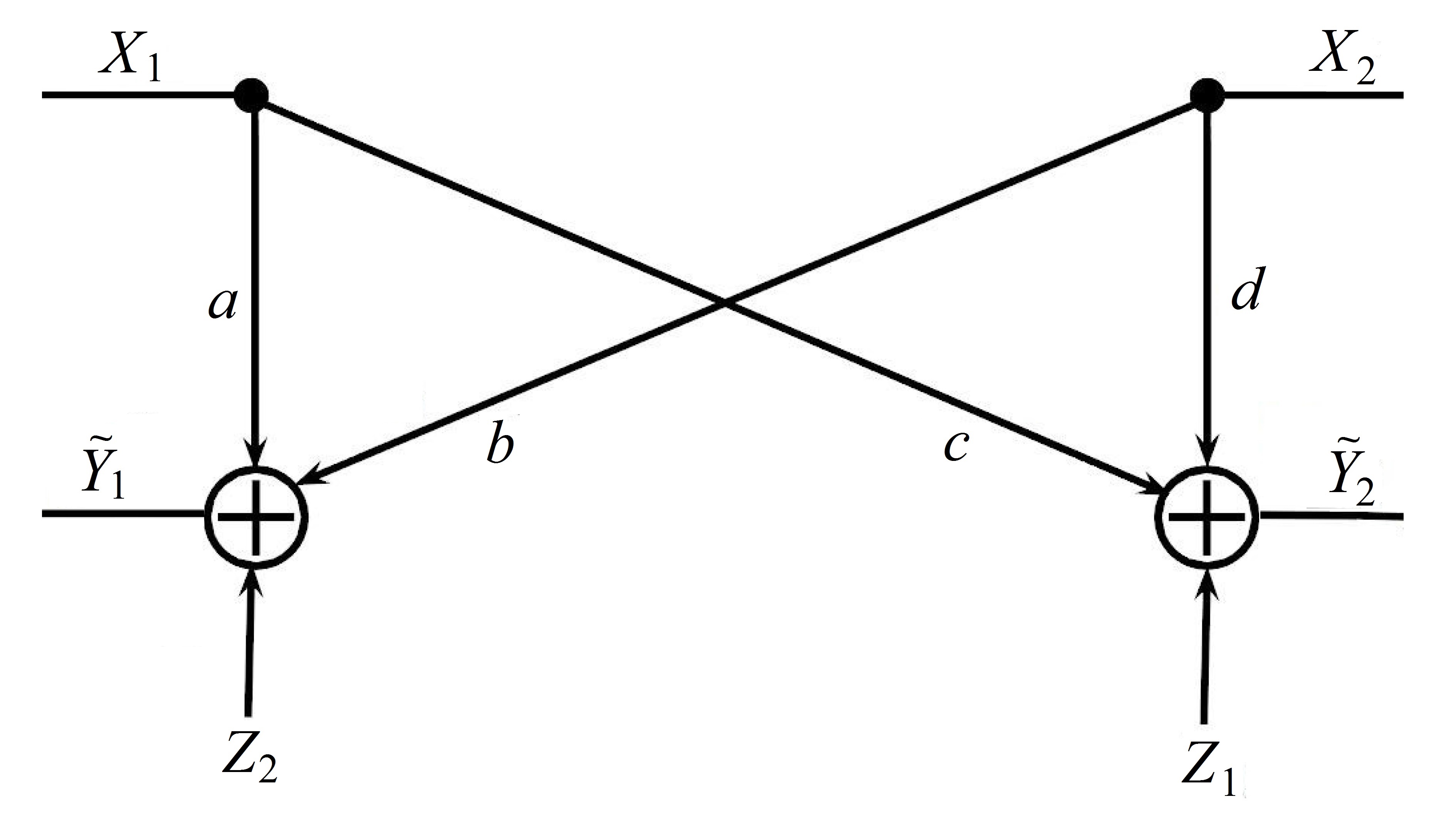}
\caption{Model of Two-Way channel}
\label{fig: channel}
\end{figure}

An analysis on two-way erasure channels was presented in \cite{khodam}. In recent years, the two-way relay channel has attracted many researchers, from both analysis \cite{relay1} and design \cite{relay3} points of view. In a two-way relay channel, it is usually assumed that there is no direct link between the two nodes and transmission is facilitated using relay nodes. Different strategies for relaying were designed for this network and articles about this type of channel compose most of the two-way channel literature. 

However, in this paper, we try to address some problems that may arise for conventional two-way channels, Fig. \ref{fig: channel}, in real world. One such problem comes from quantizing the received signals for further processing. This may make the capacity region in (\ref{r1}) invalid.  

Quantization is an inevitable part of modern communication systems. Most of signal processing operations at the receiver side are performed after the analog-to-digital conversion stage. Consider the GTWC with output quantization in Fig \ref{fig: quantizer}. 

\begin{figure}[!h]
\centering
\includegraphics[trim = 0mm 0mm 0mm 5mm,width=50mm]{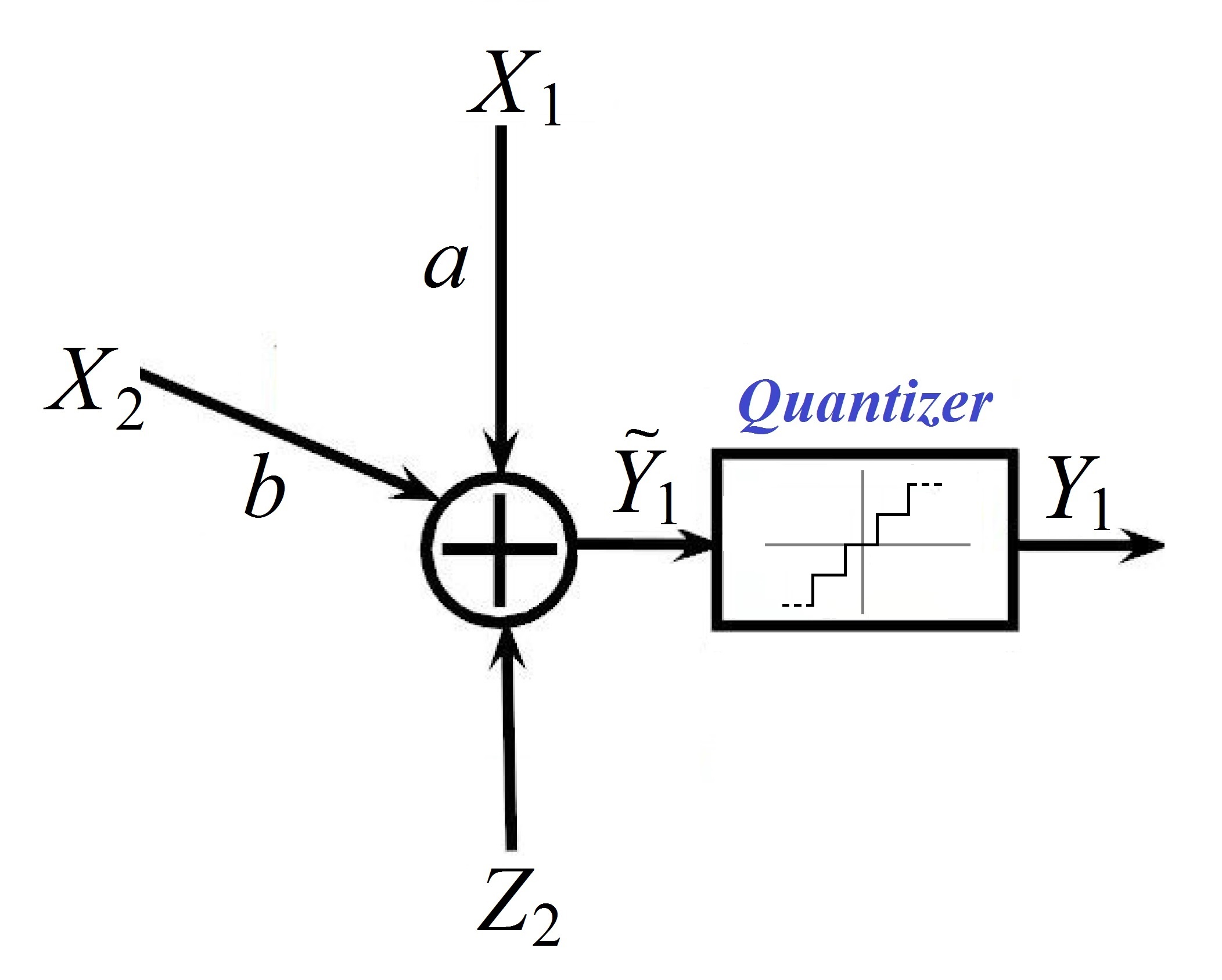}
\caption{GTWC with a saturating quantizer at the output}
\label{fig: quantizer}
\end{figure}

The system model is given by
\begin{equation}
\begin{array}{cc}
Y_1=Q(\tilde{Y}_1)=Q(aX_1+bX_2+Z_2),\\\\
Y_2=Q(\tilde{Y}_2)=Q(cX_1+dX_2+Z_1),
\end{array}
\end{equation}
where $Y_1$ and $Y_2$ are the quantized outputs and $Q(\cdot)$ is quantization function. Since quantization is a nonlinear operation, users cannot cancel the effect of self-interference anymore. Therefore, in contrast to GTWC, Gaussian inputs are not necessarily optimal.

We utilize identical quantizers with a finite number of quantization levels at both ends. The grain size of the quantizers is denoted by  $q$. The output of the quantizer can take any of the $M$ real numbers in the set $\mathcal{Y}\!=\{l_1,l_2,...,l_M \}$. In fact, $Q(y)=l_{i}$ whenever $y\in \mathcal{R}_{i}=[b_{i-1},b_{i}]$ where
\begin{IEEEeqnarray*}{clr}
b_0=-\infty\\
b_M=+\infty \\
b_i=\left(i-\frac{M}{2}\right)q, \,\,\,\,\,\,
i\in\{1,2,...,M-1\}.\IEEEyesnumber
\end{IEEEeqnarray*}

We take $\frac{P}{\sigma_{z}^{2}}$ as the measure of $\mathrm{SNR}$. In \cite{Q1,Q2}, it is shown that in a point-to-point Gaussian channel with quantized output, the capacity achieving input distribution is discrete with a finite number of mass points. In the setup of a GTWC with quantized outputs, our results confirm the supremacy of discrete inputs over Gaussian inputs at
least in the low SNR regime. As such, the majority of the paper is devoted to constellation-based transmitters.

In \cite{Q1}, it is proposed that the loss in mutual information between the input and output of a point-to-point channel due to low-precision quantization is
tolerable and even for high values of SNR ($20$ dB), $3$-bit quantizers do not
decrease the performance more than 15$\%$ compared to infinite
precision quantization. Motivated by this observation, we rely on $8$-level
($3$-bit) quantizers in our simulations unless otherwise stated.

The rest of this paper is organized as following. In section II,
performance of Gaussian inputs is shortly studied and optimum grain
size of quantizer is computed numerically for some $\mathrm{SNR}$s. In section III, for 1-dimensional and 2-dimensional scenarios, the expression for achievable rate region with constellation-based
inputs is derived. In section IV, we consider a $\theta$
degrees rotation in constellation of one of the users, and
investigate its results on rate region.

\end{section}
\begin{section}{Gaussian Inputs}

Although Gaussian inputs are not necessarily optimal for our problem, it is
still of interest to evaluate
 their performance in this model. For the sake of simplicity, we assume that all channel gains are identical and equal to one. Moreover, we set $P_{1}=P_{2}=P$. Due to symmetry, we focus on computing $R_{1}$.
According to (\ref{eqn: rate}), we need to compute $I(X_1;Y_2| X_2)$. Note that $Y_2$ is a discrete random variable. Deriving a closed form for this conditional mutual information is
unlikely. However, we can compute it numerically and find the optimum quantizer.

Fig. \ref{fig:awesome_image} demonstrates the optimum grain size of output quantizers, which maximizes the rate, for
different values of $\mathrm{SNR}$.

\begin{figure*}[!t]
\normalsize
\setcounter{equation}{10}
\begin{IEEEeqnarray*}{clr}
P(Y_2=l_k\mid X_2=x_{2,i},X_1=x_{1,j})
=P(\tilde{Y}_2\in \mathcal{R}_k\mid X_2=x_{2,i},X_1=x_{1,j})
=\int_{b_{i-1}}^{b_i}\frac{1}{\sqrt{2\pi}}e^{-\frac{(\tilde{Y}_2-x_{2,i}-x_{1,j})^2}{2}}d\tilde {Y}_2\\
=\phi (b_i-x_{2,i}-x_{1,j})- \phi (b_{i-1}-x_{2,i}-x_{1,j}) \label{eqn_1dProb}\IEEEyesnumber
\end{IEEEeqnarray*}
\setcounter{equation}{12}
\begin{equation}
H(Y_2\mid X_1=x_{1,j},X_2=x_{2,i})=- \sum_{k=1}^M P(Y_2=l_k\mid X_2=x_{2,i},X_1=x_{1,j}) \log_2 P(Y_2=l_k\mid X_2=x_{2,i},X_1=x_{1,j}) \label{eqn:_1dHy2x2x2x1x1}
\end{equation}
\hrulefill
\vspace*{4pt}
\end{figure*}

\begin{figure}[h]
\centering
\includegraphics[trim = 10mm 0mm 10mm 0mm,width=90mm]{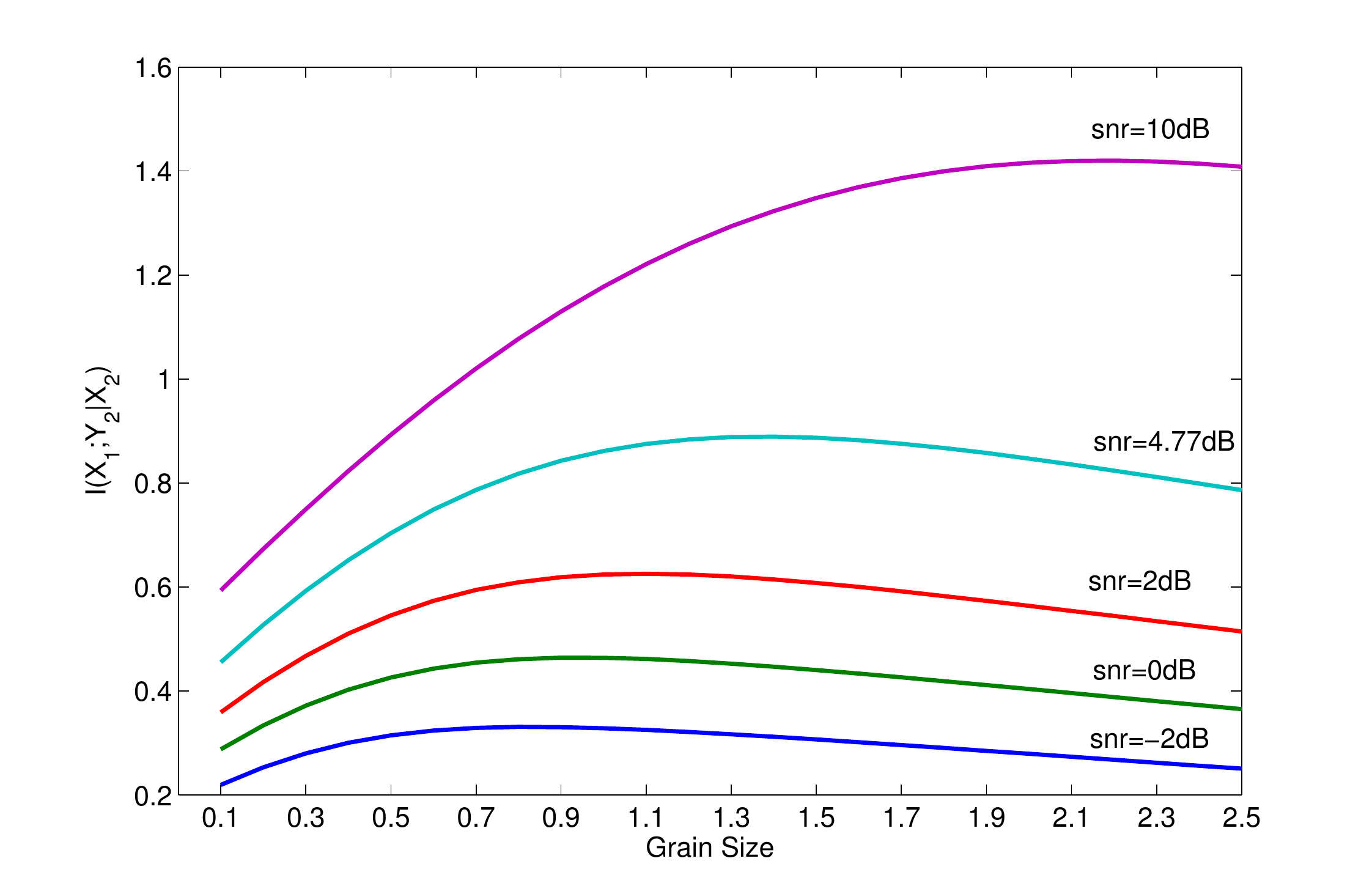}
\caption{Optimum quantizer grain size for GTWC with Gaussian inputs at different SNRs}
\label{fig:awesome_image}
\end{figure}

The following observations can be made from this
figure:

1- Low-precision quantizing does not affect performance considerably.
For example, at $\mathrm{SNR}=4.77$ dB, the best rate we can achieve is $0.89$ bits/sec/hz
with grain size $1.3$. If we do not use a quantizer,
this rate would be $1$ bits/sec/hz according to (\ref{r1}). This implies that there is about $10\%$ loss due to
3-bit quantization in contrast to the case with no quantization.

2- Fixing the value of $M$, there is only one optimum grain size. In fact, for small grain
sizes, the quantizer cannot cover the whole dynamic range of its input. On the other hand, as we
increase the grain size, the resolution decreases. This results in
loss of information as well. The reduction continues until we reach a point
in which almost the whole signal lies in one step and the amount of
$I(X_1;Y_2\mid X_2)$ converges to a certain number (e.g., $0.37814$
for $\mathrm {SNR}=4.77$ dB).

3- As $\mathrm{SNR}$ increases the dynamic range of the signal at the quantizer input
grows and the optimum grain size increases accordingly.
\end{section}

\begin{section}{Constellation-based Inputs}
Next, we evaluate the Shannon achievable region in a GTWC with constellation-based inputs. Simulation results in
Table 1 compares the values of $R_{1}$ in constellation-based GTWC with PAM signals and  GTWC with Gaussian inputs. According to this Table, at least at low $\mathrm{SNR}$ discrete input has  supremacy over Gaussian. We didn't optimize over all discrete inputs though, and just used identical 8-points PAM with different power constraint for both transmitters. For the rest of this paper, we assume that the noise power is equal to $1$, i.e.,  $\sigma_z^2=1$ and channel gains are symmetric, i.e., $a=d$ and $b=c$.

\begin{table}[!h]
    \centering
  \caption{Performance of Gaussian and Discrete Inputs in a GTWC with Output Quantization}
    \begin{tabular}{ | c ||  c| c || c | c |}
    \hline
    \multirow{2}{*}{SNR} & \multicolumn{2}{|c||}{Gaussian Inputs } & \multicolumn{2}{|c|}{Discrete Inputs (PAM)} \\ \cline{2-5}
    & $R_1$ & Opt. Grain Size & $R_1$ & Opt. Grain Size    \\ \hline
    1   & 0.46432    &0.95    &0.46972    &0.85  \\ \hline
    2   & 0.71814    &1.2    &0.72418    &1.05 \\ \hline
    3   & 0.88916    &1.4    &0.89247    &1.2 \\ \hline
    4   & 1.0162    &1.55    &1.0165        &1.4 \\ \hline
    5   & 1.116        &1.65    &1.1125        &1.5 \\ \hline
    6   & 1.1976    &1.8    &1.1911        &1.7 \\ \hline
    7   & 1.2659    &1.9    &1.2564        &1.8 \\ \hline
    \end{tabular}
\end{table}

Suppose $X_1$ and $X_2$ are generated uniformly over
finite constellations $\mathcal{X}_1$ and $\mathcal{X}_2$ with
cardinality $K_1$ and $K_2$, respectively, i.e.,
$\mathcal{X}_1=\{x_{1,1},x_{1,2},...,x_{1,K_1}\}$ and
$\mathcal{X}_2=\{x_{2,1},x_{2,2},...,x_{2,K_2}\}$. One may
express $I(X_1;Y_2| X_2)$ as
\setcounter{equation}{5}
\begin{IEEEeqnarray}{rCl}
I(X_1;Y_2\mid X_2)=H(Y_2\mid X_2)-H(Y_2\mid X_1,X_2) \label{Eqn: Cond Mut}.
\end{IEEEeqnarray}

For $I(X_2;Y_1\mid X_1)$ we will have exactly the same arguments as (\ref{Eqn: Cond Mut}) and just need to exchange the indices. 

We study both 1-dimension and 2-dimension scenarios in the following subsections.

\begin{subsection}{1-Dimensional Constellations}
In this subsection, we consider a constellation with points along one axis. For such constellation, 
$H(Y_2\mid X_2)$ in (\ref{Eqn: Cond Mut}) has the following form:
\begin{IEEEeqnarray}{rCl}
H(Y_2\mid X_2)=\frac{1}{K_2} \sum_{i=1}^{K_2} H(Y_2\mid X_2=x_{2,i}) \label{eqn:_Hy2x2}
\end{IEEEeqnarray}
and
\begin{IEEEeqnarray*}{clr}
H(Y_2\mid X_2=x_{2,i})=  \\
-\sum_{k=1}^M P(Y_2=l_k\mid X_2=x_{2,i})\log _2P(Y_2=l_k\mid X_2=x_{2,i}). \label{eqn:_1dHy2x2x2}\\
\IEEEyesnumber
\end{IEEEeqnarray*}
On the other hand,
\begin{IEEEeqnarray*}{clr}
P(Y_2=l_k\mid X_2=x_{2,i})= \\
\frac{1}{K_1} \sum_{j=1}^{K_1} P(Y_2=l_k\mid X_2=x_{2,i},X_1=x_{1,j}) \label{eqn:_1dPy2x2}
\IEEEyesnumber
\end{IEEEeqnarray*}

We need to discuss about (\ref{eqn:_1dPy2x2}). Note that $Y_2$ is a quantized
version of $\tilde{Y}_2$ and the probability density function of $\tilde{Y}_2$ is
\begin{equation}
 f(\tilde{Y}_2\mid X_2=x_{2,i},X_1=x_{1,j})=\frac{1}{\sqrt{2\pi}}e^{-\frac{(\tilde{Y}_2-dx_{2,i}-cx_{1,j})^2}{2}} \label{eqn:_1dPy2x2x1}.
\end{equation}

This leads us to (\ref{eqn_1dProb}) where $\phi(\cdot)$ is the cumulative distribution function of a standard
Gaussian random variable.
\setcounter{equation}{11}
As for $H(Y_2\mid X_1,X_2)$,
\begin{IEEEeqnarray*}{clr}
H(Y_2\mid X_1,X_2)\!=\!\frac{1}{K_1}\frac{1}{K_2}\sum_{j=1}^{K_1}\!
\sum_{i=1}^{K_2} H(Y_2\!\mid\! X_1\!=\!x_{1,j},X_2\!=\!x_{2,i}).  \label{eqn:_Hy2x2x1} \\
\IEEEyesnumber
\end{IEEEeqnarray*}
\setcounter{equation}{13}
Similarly, $H(Y_2\mid X_1=x_{1,j},X_2=x_{2,i})$ can be written as (\ref {eqn:_1dHy2x2x2x1x1}).
\end{subsection}
\begin{subsection}{2-Dimensional Constellations}
Next, we consider 2-Dimensional Constellations. The
ambient noise at both ends is modeled as circularly symmetric complex Gaussian noise with unit variance.
\begin{figure*}[!t]
\normalsize
\setcounter{equation}{13}

\begin{IEEEeqnarray}{clr}
H(Y_2\mid X_2=x_{2,i})= -\sum_{m=1}^M \sum_{n=1}^N P(Y_2=l_{mn}\mid X_2=x_{2,i})\log _2P(Y_2=l_{mn}\mid X_2=x_{2,i}) \label{eqn:_2dHy2x2x2}
\end{IEEEeqnarray}

\begin{IEEEeqnarray}{clr}
P(Y_2=l_{mn}\mid X_2=x_{2,i})=\frac{1}{K_1} \sum_{j=1}^{K_1} P(Y_2=l_{mn}\mid X_2=x_{2,i},X_1=x_{1,j})
\end{IEEEeqnarray}

\begin{IEEEeqnarray}{clr}
f(\tilde{Y}_2\mid X_2=x_{2,i},X_1=x_{1,j})=\frac{1}{\pi}e^{-\mid \tilde{Y}_2-x_{2,i}-x_{1,j}\mid^2}
\end{IEEEeqnarray}

\begin{IEEEeqnarray*}{clr}
\label{eqn_khafan}
P(Y_2=l_{mn}\mid X_2=x_{2,i},X_1=x_{1,j})=P(\tilde{Y}_2\in \mathcal{R}_{mn}\mid X_2=x_{2,i},X_1=x_{1,j})
=\int_{b_{m-1}}^{b_m} \int_{d_{n-1}}^{d_n} \frac{1}{\pi}e^{-\mid \tilde{Y}_2-x_{2,i}-x_{1,j}\mid^2}d\tilde {Y}_2^{(1)}d\tilde {Y}_2^{(2)}\\
\!=\!\Big[\phi (\sqrt{2}(b_m-x_{2,i}^{(1)}-x_{1,j}^{(1)}))\!-\!\phi (\sqrt{2}(b_{m-1}-x_{2,i}^{(1)}-x_{1,j}^{(1)}))\Big]\Big[\phi (\sqrt{2}(d_n-x_{2,i}^{(2)}-x_{1,j}^{(2)}))\!-\!\phi(\sqrt{2}(d_{n-1}-x_{2,i}^{(2)}-x_{1,j}^{(2)}))\Big] \IEEEyesnumber
\end{IEEEeqnarray*}

\begin{equation}
H(Y_2\mid X_1=x_{1,j},X_2=x_{2,i})=-\sum_{m=1}^M \sum_{n=1}^N P(Y_2=l_{mn}\mid X_2=x_{2,i},X_1=x_{1,j}) \log_2 P(Y_2=l_{mn}\mid X_2=x_{2,i},X_1=x_{1,j}) \label{eqn:_2dHy2x2x2x1x1}
\end{equation}
\hrulefill
\vspace*{4pt}
\end{figure*}

We need to perform 2-dimensional quantization at outputs. Quantization is performed independently on each dimension. 
Due to uniform quantization, the quantizer regions, $\mathcal{R}_{mn}$, will be rectangular with horizontal boundaries 
$b_{m-1}$ and $b_m$ and vertical boundaries $d_{n-1}$  and $d_n$. Let us denote the quantization levels by $l_{mn}$. 
Assume that the quantizers have $M$ horizontal and $N$ vertical levels. If $y_i \in \mathcal{R}_{mn}$ then $Q(y_i)=l_{mn}$ 
(for $i=1,2$). Basically, expressions for obtaining conditional mutual information in 2-dimensional case can be derived in 
an almost similar manner to 1-dimensional problem. However, they are slightly different. Equations (\ref{eqn:_Hy2x2}) and 
(\ref{eqn:_Hy2x2x1})  remain unchanged. However, equations (\ref{eqn:_1dHy2x2x2}) to (\ref{eqn_1dProb}) 
and (\ref{eqn:_1dHy2x2x2x1x1}) change to equations (\ref{eqn:_2dHy2x2x2}) to (\ref{eqn:_2dHy2x2x2x1x1}) where $T^{(1)}$ 
and $T^{(2)}$ denote components of variable $T$, $T=T^{(1)}+\sqrt{-1}T^{(2)}$. Note that we need to rely on numerical computations.

In the next section, the rate region will be sketched for 4-PAM and QPSK at some $\mathrm{SNR}$s.
\end{subsection}
\end{section}

\begin{section}{Rotation of Constellation}
In this section we extend the concept of Uniquely Decodable (UD) alphabet pairs proposed in \cite{hr}. 

For given constellations $\mathcal{X}_1$ and $\mathcal{X}_2$, $\mathcal{X}_{sum1}$ and $\mathcal{X}_{sum2}$ are defined as follow (given $a=d$ and $b=c$):
\pagebreak
\begin{IEEEeqnarray*}{clr}
\mathcal{X}_{sum1}=\{Q(ax_1+bx_2) \mid \forall x_1 \in \mathcal{X}_1, x_2 \in \mathcal{X}_2\} \\
\mathcal{X}_{sum2}=\{Q(bx_1+ax_2) \mid \forall x_1 \in \mathcal{X}_1, x_2 \in \mathcal{X}_2\}
\end{IEEEeqnarray*}
In fact, $\mathcal{X}_{sum1}$ and $\mathcal{X}_{sum2}$ denote the quantized version of received constellations at each receiver. Given the mappings $\psi_1: \mathcal{X}_1 \times \mathcal{X}_2\mapsto\mathcal{X}_{sum1}$ and $\psi_2: \mathcal{X}_1 \times \mathcal{X}_2\mapsto\mathcal{X}_{sum2}$, we call the pair $(\mathcal{X}_1,\mathcal{X}_2)$ to be a UD pair if $\psi_1$ and $\psi_2$ are one-to-one mappings.

If the pair $(\mathcal{X}_1,\mathcal{X}_2)$ is UD, probability of error in decoding the received signal decreases and information can be transmitted through the channel at higher rates.

A simple way to achieve such UD pairs is to rotate the constellation of one user, i.e, $\mathcal{X}_2=\mathcal{X}_1 e^{j\theta}$. As such, we let $K_1=K_2=K$. Our goal is to find an angle of rotation that maximally enlarges  the Shannon achievable region. Let us denote such an angle by $\theta^*$. Numerical simulations show that the rotation of one constellation enlarges the achievable region and in some cases, results in a rectangular region. According to the definition of UD pairs, it is clear that in some cases, constellation rotation does not help us in reaching our goal, i.e., $\theta^*=0$, specially for quantizers with large grain size. In fact, the optimum value of $\theta$ depends on the structure of the quantizer. Generally, for 1-dimensional constellations, $\theta^{*}=90$ for most of the cases. For 2-dimensional constellation, by increasing the number of constellation points, the optimum angle decreases.

For a UD constellation pair, both $\mathcal{X}_{sum1}$ and $\mathcal{X}_{sum2}$ have $K^2$ elements. As $\mathrm{SNR}$ increases, sum rate converges to $\log_2K^2=2\log_2K$, which is the maximum achievable sum-rate for a channel with $K$-point constellations at inputs.

It is necessary to mention that if we do not quantize the output, rotation of constellation does not help in enlarging the rate region, because the receiver knows the constellation. Therefore its rotation does not provide any further information. However, since the quantizer does not operate linearly, its output is not completely clear for the receiver. From a mathematical point of view, (\ref{eqn_khafan}) without quantization is an integral from $-\infty$ to $+\infty$ and rotation, which is equivalent to changing the mean value of the random variable $\tilde{Y}_2$, does not have any effect on the results. But, because we are integrating on a bounded interval, location of the mean value of $\tilde{Y}_2$ is important.


\subsection{Applying Rotation method to QPSK and 4-PAM}
In this section the effect of rotation of constellation is studied for some practical constellation choices.  In all of the results of this section, we assume all channel gains are equal to $1$, and grain size of the quantizer is also equal to $1$.

We first  apply this method to a 4-PAM constellation.  As it is illustrated in Fig. \ref{fig:awesome_image11}, rotation enlarges the achievable rate region considerably, specially for higher values of $\mathrm{SNR}$. Without rotation we have only one dimension in transmission. Through applying rotation, we are adding another dimension which decreases the effect of self-interference.

Fig. \ref{fig:awesome_image12} shows the results of rotation of one QPSK constellation. Here, we can see the advantage of rotation as well. In a moderate $\mathrm{SNR}$ ($10$ dB) we can almost achieve 2 bits/sec/hz for each user which is the maximum achievable rate when we use this particular constellation.

We can also compare the performance of these two constellations. For all amounts of $\mathrm {SNR}$, QPSK works better than PAM, as it was expected. But, for PAM, improvement obtained by rotation of constellation is much larger than QPSK. This is due to the orthogonality ($\theta^*=90$ for PAM) caused by rotation for 1-dimensional constellations.

\begin{figure}[!b]
\centering
\includegraphics[trim = 10mm 0mm 10mm 0mm,width=80mm]{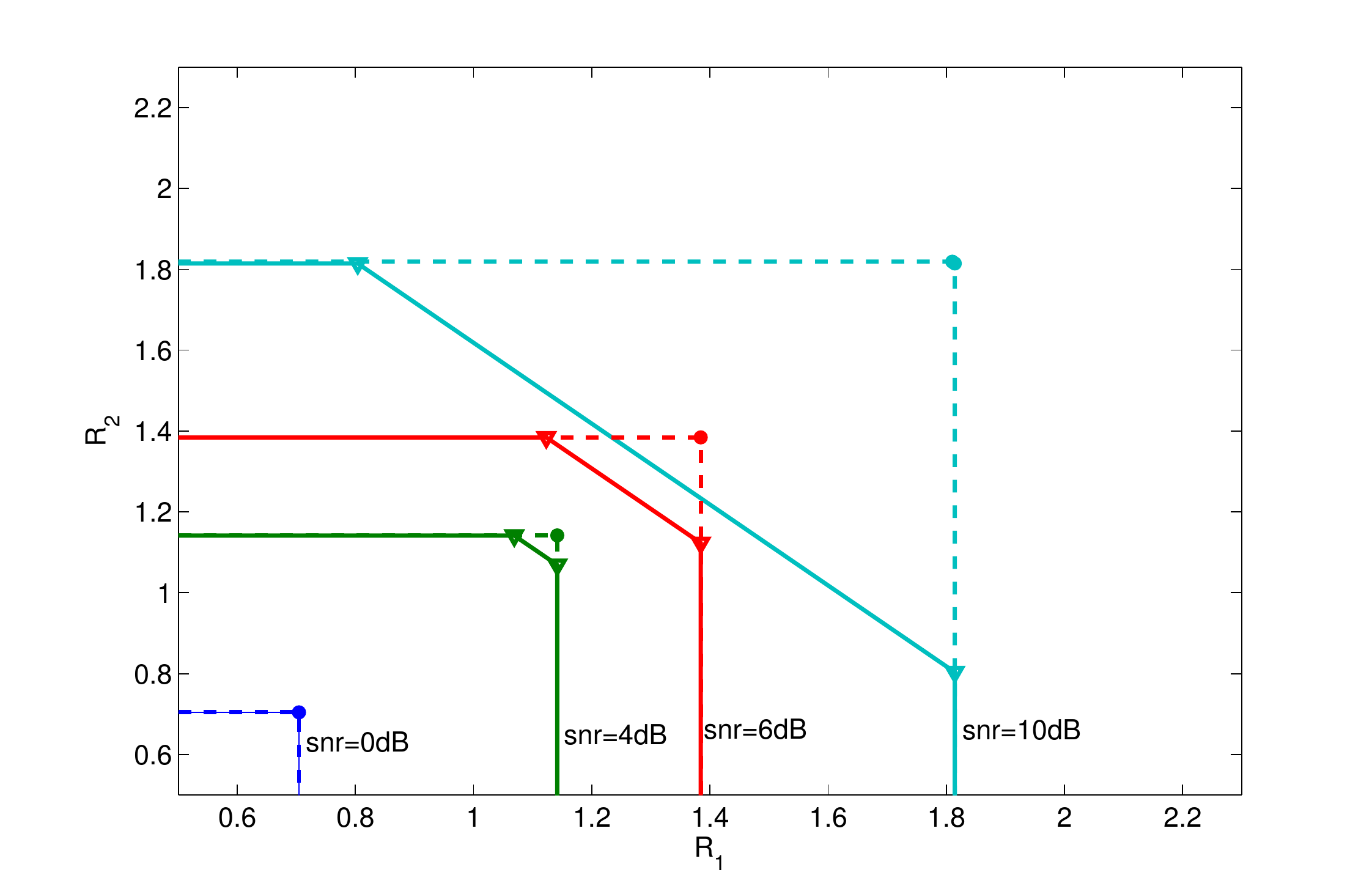}
\caption{Result of Rotation of Constellation for 4-PAM at different SNRs- Dashed: with rotation, Solid: without rotation}
\label{fig:awesome_image11}
\end{figure}

\begin{figure}[!t]
\includegraphics[trim = 10mm 10mm 10mm 10mm,width=80mm]{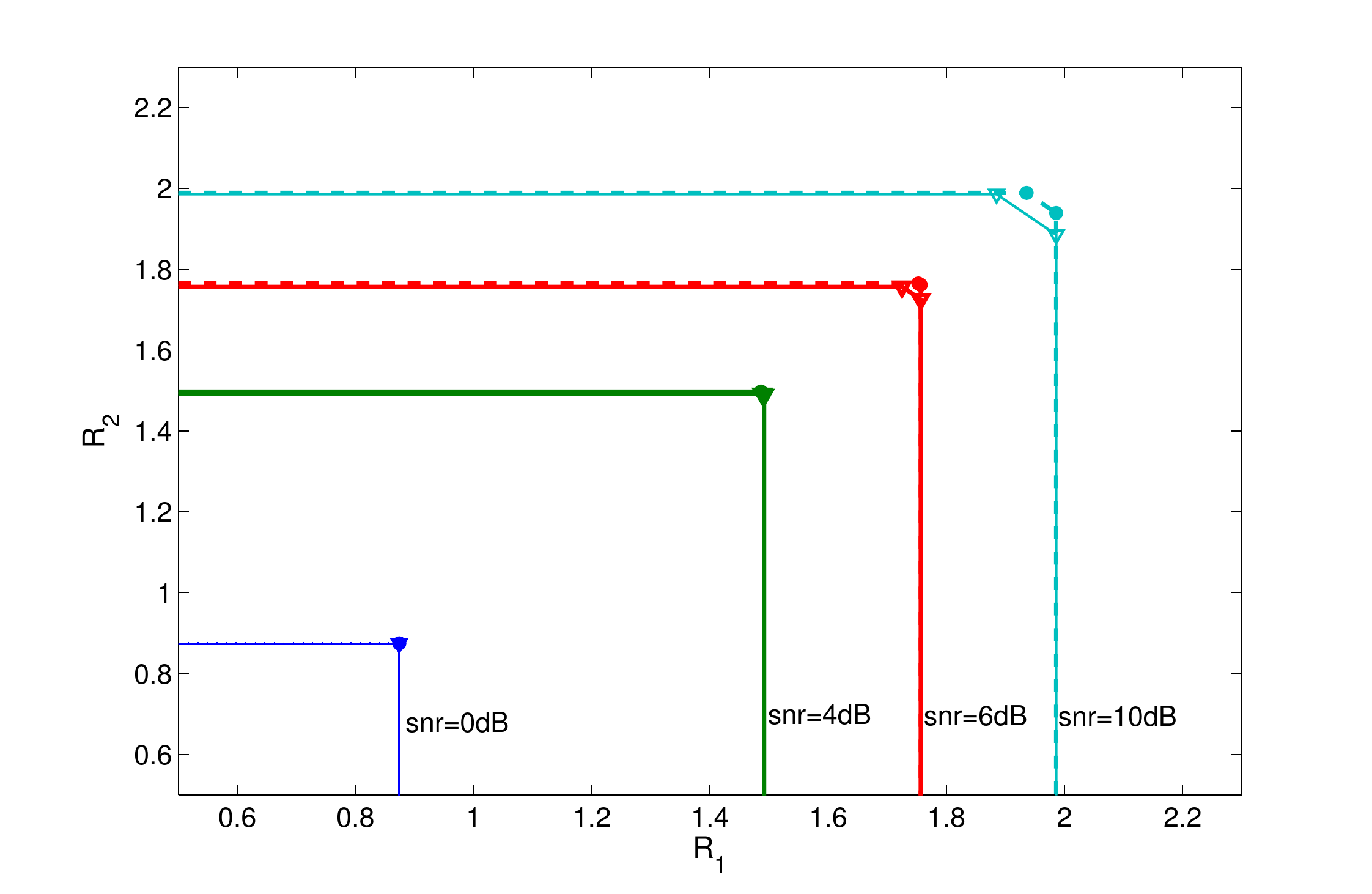}
\caption{Result of Rotation of Constellation for QPSK at different SNRs- Dashed: with rotation, Solid: without rotation}
\label{fig:awesome_image12}
\end{figure}
\end{section}
\vspace{-0cm}
\begin{section}{Conclusion and Future Work}
\vspace{-.1cm}
We considered the effect of presence of uniform saturating quantizers at the receivers of Gaussian two-way channel. As the input of this channel, some discrete constellation was used which had better performance than Gaussian inputs, despite their simplicity.  A formulation for achievable rate region of this channel using discrete inputs was derived and it was shown that rotation of the constellation can enlarge the rate region and help us to transmit the messages with higher rate.

Proving the optimality of discrete input can be an interesting problem for this channel. A possible way to do so is by showing that the presence of quantizers at the outputs of GTWC can implicitly impose a peak power constraint on the input signals. This approach was also employed in \cite{Q1}. If this is shown, the results of  \cite{smith} can be used to prove the optimality.   
\end{section}

\end{document}